 \documentclass[final,1p,times,twocolumn,authoryear]{elsarticle}

\usepackage{amssymb}

\usepackage{textcomp}
\DeclareTextSymbolDefault{\degreesymbol}{TS1} 
\DeclareTextSymbol{\degreesymbol}{TS1}{176} 
\DeclareRobustCommand{\textdegree}{\ifmmode\mbox{\degreesymbol}\else\degreesymbol\fi}

\usepackage{graphicx}	
\usepackage{amsmath}	
\usepackage{amssymb}	
\usepackage{rotating}
\usepackage{lscape}
\usepackage{comment}
\usepackage[utf8]{inputenc}
\usepackage[dvipsnames]{xcolor}
\usepackage{siunitx}
\usepackage{verbatim}
\usepackage[english]{babel}
\usepackage[T1]{fontenc}
\usepackage{ae,aecompl}
\usepackage{rotating}
\usepackage{textcomp}
\usepackage{lipsum}
\usepackage{url}
\usepackage[normalem]{ulem}
\usepackage{soul}

\begin{document}

\begin{frontmatter}



\title{Searching for misaligned active galactic nuclei among blazar candidates in the Fourth {\it Fermi}-LAT catalog.}

 \author[label1,label2]{Graziano Chiaro}
 \address[label1]{Istituto di Fisica Cosmica e Astrofisica Spaziale  IASF, INAF Via A. Corti 12,  20133 Milano  IT}  
 \address[label2]{Consorzio Interuniversitario per la Fisica Spaziale  CIFS ,Via Pietro Giuria, 1, 10125 Torino IT}  
 \address[label3]{Lab. de Instrumenta\c{c}\~ao e F\'{i}sica Experimental de Part\'{i}culas LIP, Av. Gama Pinto 2, Lisboa, PT}
\address[label4]{IPARCOS and Department of EMFTEL, Universidad Complutense de Madrid, Av. Séneca, 2, 28040 Madrid, ES}

\author[label3]{Giovanni La Mura}
\author[label4]{Alberto Dom{\'i}nguez}
\author[label1]{Susanna Bisogni}

\begin{abstract}
Radio-loud sources with blazar-like properties, but having a jet that does not directly point in the direction of the observer are among the most interesting classes of $\gamma$-ray emitters. These sources are known as Misaligned Active Galactic Nuclei (MAGN). Understanding MAGN properties is useful to improve the knowledge of blazar energetics. We searched for new MAGN candidates among the remaining blazars of uncertain type  detected by the {\it Fermi} Large Area Telescope (LAT) using a methodology based on characterizing their radio morphology. We identified seven candidates associated with $\gamma$-ray sources. Their features are consistent
with a source with a misaligned relativistic jet consistent with the definition of MAGN.
\end{abstract}




\end{frontmatter}


\section{Introduction}
\label{<1>}

Launched by NASA on 2008 June 11 the {\it Fermi} satellite has been observing $\gamma$-ray emitters with the Large Area Telescope (LAT) the primary instrument onboard \citep{lat}. 
LAT is an imaging, high-energy $\gamma$-ray telescope, covering the energy range from below 100 MeV to more than 300 GeV. The Fourth {\it Fermi}-LAT Source Catalog (4FGL) reports $\gamma$-ray data collected by LAT in the first eight years of operation and lists 5065 sources \citep{4fgl}.
Out of these, 3131 sources are blazars, which are divided into 1116 BL Lacertae (BL Lac), 686 Flat Spectrum Radio Quasar (FSRQ) and 1329 blazars candidates of uncertain type (BCU). {\bf{These sources are also}} reported in the Fourth Catalog of Active Galactic Nuclei \citep[4LAC,][]{4lac}.\\
Gamma-ray emitters that host an Active Galactic Nucleus (AGN) can be classified according to the orientation of  their relativistic jet relative to the line of sight of the observer, as described in the unified model \citep{urry}. In that model, blazars are active galaxies whose jets are seen at a small viewing angle.  
Another class of AGNs, known as Misaligned Active Galactic Nuclei \citep[MAGN,][]{magn} show a jet of radiation pointed at larger angles to the viewer than blazars. 
The radio-loud MAGNs are further classified into two main classes according to the distance of the brightest point from the central core \citep{urry,magn}, namely the
edge-darkened Fanaroff-Riley I radio galaxies (FR-I) and the edge-brightened Fanaroff-Riley
II radio galaxies (FR-II) \citep{fan}. 
Steep Spectrum Radio Quasars (SSRQs), whose jet angles are smaller than those of radio galaxies but larger than those of blazars, are often also considered as MAGNs \citep{ori}. We do not include SSRQs in the present analysis.
In \citet{sar} and in \citet{chia} the authors gave several reasons to support the view that FR-I and FR-II are 
the parent population of blazars. This interpretation is also supported by the similarity in the radio-to-optical and optical spectral indices of the two classes of sources. From this relationship with blazars, studying MAGNs remains  of fundamental importance for the understanding of the AGNs population as a whole, their radiation mechanisms, and their galaxy structure as well as 
the disparity of detection between more powerful and weaker $\gamma$-ray sources \citep{arm,ghi,tor}. There are 36 $\gamma$-ray sources classified as MAGN in 4LAC, however the number could be greater if new candidates whose properties are closely related with characteristic MAGN features could be found among the remaining  blazars of uncertain type (BCUs). It is the aim of this study to search for such sources.\\
The paper is organized as follows: Sec.~\ref{<2>} describes our selection criteria and radio data; in Sec.~\ref{<3>} we discuss the results and, finally, Sec.~\ref{<4>} presents a brief conclusion.

\section{Sample selection and data}
\label{<2>}
In this section, we select potential MAGNs from BCUs in 4LAC based on the $\gamma$-ray properties of known MAGNs listed in the catalog. Then, we search for these sources in archival radio data and characterize their extension. This procedure allows us to find the best candidates.

\subsection{Gamma-ray selection}
We began this study by searching for the most probable MAGN candidates from the BCU population in the 4LAC catalog, using the fact that their properties cover only a fraction of the parameter space spanned by blazar spectra. Using a power-law with index $\Gamma$\ to represent the spectra in the form of ${\rm d}N/{\rm d}E \propto E^{-\Gamma}$, it is found that MAGNs lie within the range $1.6 < \Gamma < 2.7$, while blazars generally have $\Gamma > 1.2$ 
Furthermore we considered the Variability Index (VI). This parameter is defined as twice the sum of the log-likelihood difference between the flux fitted in each time interval and the average flux over the full catalog interval and it represents the statistical significance of the variability of a source, with respect to its averaged spectral features.It is a key property of blazars depending on the energy band considered \citep{var1,var2,var3} and explained as emitting electrons in a leptonic scenario of different energies and thus different acceleration/cooling times in the distinct bands. Known MAGNs have $2.8 < {\rm VI} < 30$. This range indicates no significant variability, in contrast to blazars, which are often highly variable, with more of $40\%$ of them having ${\rm VI} > 72$ \citep{4lac}.We applied these two criteria to the BCU sources selecting 1062 candidates. 

\subsection{Radio data}
A key parameter for a rigorous misaligned AGN classification is the orientation angle of the relativistic jet relative to the line of sight. Yet this parameter is difficult to determine through observations and most of the methods in the literature provide uncertain measurements or limits \citep{landt,giov}.
Approximate estimates of the inclination can be obtained from measurements of the projected size of the extended emission, if we can make assumptions on the intrinsic size of the sources. Unfortunately, the observed differences in the extension of powerful radio galaxies seen at very large angles only allow to place weak constraints on the limiting values of the inclination angle. Furthermore, it is not obvious what angle defines the frontier between a blazar and a MAGN. 
Another method to measure the orientation angle of an AGN jet relies on the determination of the [O\small{~III}] $\lambda 5007$ emission line equivalent width. Since the line radiation is emitted almost isotropically from an extended ionized region, while the underlying AGN continuum is prone to obscuration, most likely close to the plane of the accretion disk, the ratio of the corresponding intensities is expected to depend on the inclination of the source \citep{Risaliti2011,Bisogni2017,Bisogni2019}. In this case, the main factors of indetermination are the uncertainty in the relation between the AGN continuum and the emission line, which depends on the covering factor and the distribution of the ionized gas, and the possible differences among the jet orientation and the distribution of the obscuring material. However, searching for the sources in our sample in the spectroscopic databases of the Sloan Digital Sky Survey \citep[SDSS,][]{sdss1, sdss2}, of the 6dF Galaxy Redshift Survey \citep[6dFGS,][]{jon1, jon2}, as well as in dedicated spectroscopic campaigns of 4FGL BCU lists \citep{demen}, only $7$ spectra could be found. These were BL Lac or elliptical galaxy spectra, with no indications of emission line features. By definition, these objects are not suitable for the [O\small{~III}] $\lambda 5007$ equivalent width method.
Therefore, we looked for other alternatives based on radio data. We searched for radio counterparts in the NRAO VLA Sky Survey (NVSS)\footnote{{\tt http://www.cv.nrao.edu/nvss/}}, in the Sydney University Molonglo Sky Survey (SUMSS) \footnote{http://www.astrop.physics.usyd.edu.au/sumsscat/} , and in the Faint Images of the Radio Sky at Twenty-cm (FIRST)\footnote{{http://sundog.stsci.edu/}}. Since a MAGN defining feature is the observation of a radiative jet not pointing close to the line of sight, MAGNs will appear as sources with an extended morphology, as opposed to the compact core-dominated shape of blazars. Unfortunately, the angular resolution of the surveys that cover the largest portions of sky, namely NVSS and SUMSS, is of the order of 30 arcsec or worse, and it is not suitable to assess the actual morphology of the selected targets. However, by comparing the NVSS images of objects that were also observed at the resolution of 5 arcsec in the FIRST survey, we are able to verify that the presence of extended structures in the high resolution observations reflects into asymmetric or elongated shapes in the low resolution images. Indeed, angular sizes significantly larger than the survey beam size can actually correspond to objects with a confidently detectable extended radio component. By selecting objects with a major axis larger than 45 arcsec from the NVSS and objects with a corresponding axis larger than 70 arcsec from the SUMSS, to account for its lower image quality, we are able to find sources with the strongest evidence of an extended radio emission component. We restricted the list of candidates only to sources with a strong indication of extended contributions in both data sets 
considering the ratio (MajAxis) / ( MinAxis) > 1.5 
consistent with an extended structure. According to the previous criteria, first, relative to the NVSS/FIRST selection and second, relative to the SUMSS selection, there are 25 and 23 $\gamma$-ray sources, respectively. These 48 candidates represent the master sample of our study and are listed in Table~\ref{table1}. We subsequently inspected the radio images of these candidates, in order to separate those with a confident extended structure from objects where the radio morphology was affected by noisy patterns or by the overlapped contribution of nearby sources. In this way, we obtained another list of seven objects with evidence of extended emission, collected in Table~\ref{table2}, five of which have publicly available optical spectra.

 \begin{table}
\caption{The master sample. The table columns represent, respectively, the 4FGL source name, the name of the low-energy counterpart associated in 4FGL, the radio coordinates and the angular size of the major axis expressed in arcmin, as measured by the relevant radio survey.}
\label{table1}
\begin{footnotesize}
\begin{tabular}{lcccc}
\hline 
Source Name	&	assoc 4FGL	&	RA radio	&	DEC radio	&	MajAxis  \\
\hline									
4FGL J0011.8$-$3142	&	SUMSS J001141$-$314220	&	2.923	&	$-$31.705	&	1.42	\\
4FGL J0015.9+2440	&	GB6 J0016+2440	&	4.015	&	24.670	&	2.7	\\
4FGL J0119.6+4158	&	2MASX J01200274+4200139	&	20.008	&	42.002	&	0.91	\\
4FGL J0119.9+4053	&	CRATES J012018+405314	&	19.949	&	40.905	&	1.42	\\
4FGL J0143.5$-$3156	&	PKS 0140$-$322	&	25.966	&	$-$31.901	&	0.50	\\
4FGL J0201.1$-$4347	&	GALEXASC J020110.83$-$434654.8	&	30.294	&	$-$43.781	&	1.19	\\
4FGL J0228.2$-$3102	&	PMN J0228$-$3102	&	37.054	&	$-$31.044	&	1.42	\\
4FGL J0233.5+0654	&	2MASS J02334098+0656114	&	38.377	&	6.923	&	0.76	\\
4FGL J0240.8$-$3401	&	NVSS J024047$-$340018	&	40.198	&	$-$34.005	&	1.28	\\
4FGL J0301.4$-$3124	&	AT20G J030116$-$312615	&	45.318	&	$-$31.438	&	1.29	\\
4FGL J0325.3+3332	&	2MASX J03251760+3332435	&	51.323	&	33.545	&	2.11	\\
4FGL J0338.9$-$2848	&	NVSS J033859$-$284619	&	54.747	&	$-$28.799	&	2.28	\\
4FGL J0342.8$-$3007	&	PKS 0340$-$302	&	55.667	&	$-$30.132	&	1.42	\\
4FGL J0345.5$-$3301	&	2MASS J03453818$-$3256462	&	56.373	&	$-$32.934	&	1.42	\\
4FGL J0348.8+4610	&	B3 0345+460	&	57.327	&	46.166	&	0.93	\\
4FGL J0507.9+4647	&	TXS 0503+466	&	76.848	&	46.761	&	1.34	\\
4FGL J0536.0$-$2754	&	PMN J0535$-$2751	&	83.965	&	$-$27.865	&	1.02	\\
4FGL J0551.8$-$3517	&	NVSS J055142$-$351527	&	88.027	&	$-$35.296	&	1.35	\\
4FGL J0602.7$-$0007	&	PMN J0602$-$0004	&	90.679	&	$-$0.074	&	1.13	\\
4FGL J0632.0$-$1032	&	TXS 0629$-$105	&	97.969	&	$-$10.564	&	1.33	\\
4FGL J0637.4$-$3537	&	WISE J063746.40$-$353648.3	&	99.440	&	$-$35.615	&	1.28	\\
4FGL J0828.6$-$0747	&	NVSS J082854$-$074854	&	127.227	&	$-$7.815	&	1.08	\\
4FGL J0829.7$-$5856	&	PMN J0829$-$5856	&	127.476	&	$-$58.959	&	1.20	\\
4FGL J0928.2$-$3048	&	PKS 0926$-$306	&	142.141	&	$-$30.828	&	1.42	\\
4FGL J0929.3$-$2414	&	NVSS J092928$-$241632	&	142.366	&	$-$24.275	&	1.3	\\
4FGL J1008.8$-$3139	&	PKS 1006$-$313	&	152.210	&	$-$31.651	&	1.42	\\
4FGL J1111.8+4858	&	1RXS J111157.8+485725	&	167.991	&	48.948	&	1.01	\\
4FGL J1208.4+6121	&	RGB J1208+613	&	182.156	&	61.355	&	1.44	\\
4FGL J1215.8$-$3732	&	PMN J1215$-$3734	&	184.008	&	$-$37.570	&	1.28	\\
4FGL J1223.6$-$3032	&	NVSS J122337$-$303246	&	185.905	&	$-$30.546	&	0.96	\\
4FGL J1232.5$-$3720	&	NVSS J123235$-$372051	&	188.149	&	$-$37.347	&	1.28	\\
4FGL J1344.4$-$3656	&	PKS 1341$-$366	&	206.098	&	$-$36.941	&	1.28	\\
4FGL J1455.4$-$3654	&	PKS 1452$-$367	&	223.787	&	$-$36.925	&	1.48	\\
4FGL J1457.8$-$4642	&	PMN J1457$-$4642	&	224.426	&	$-$46.701	&	2.22	\\
4FGL J1507.3$-$3710	&	NVSS J150720$-$370903	&	226.838	&	$-$37.205	&	1.28	\\
4FGL J1523.2$-$3941	&	2MASS J15234352$-$3936318	&	230.887	&	$-$39.611	&	1.07	\\
4FGL J1530.5$-$3026	&	NVSS J153041$-$302559	&	232.671	&	$-$30.432	&	1.42	\\
4FGL J1549.8$-$3044	&	NVSS J154946$-$304501	&	237.444	&	$-$30.750	&	1.44	\\
4FGL J1628.3$-$3343	&	NVSS J162819$-$334342	&	247.082	&	$-$33.728	&	1.42	\\
4FGL J1643.7+3317	&	NVSS J164339+331640	&	250.916	&	33.277	&	1.22	\\
4FGL J1741.9+2555	&	NVSS J174147+255443	&	265.473	&	25.949	&	0.90	\\
4FGL J1742.4$-$1518	&	PMN J1742$-$1517	&	265.548	&	$-$15.291	&	2.10	\\
4FGL J2002.6+6302	&	1RXS J200245.4+630226	&	300.687	&	63.041	&	0.99	\\
4FGL J2046.6$-$1012	&	PMN J2046$-$1010	&	311.727	&	$-$10.178	&	1.58	\\
4FGL J2049.0+1647	&	NVSS J204902+164727	&	312.260	&	16.790	&	0.95	\\
4FGL J2229.1+2254	&	NVSS J222913+225511	&	337.307	&	22.919	&	0.75	\\
4FGL J2309.7$-$3632	&	1RXS J230940.6$-$363241	&	347.418	&	$-$36.541	&	1.29	\\
\hline
\end{tabular}
\end{footnotesize}
\end{table}

\begin{figure}
\begin{center}
\includegraphics[width=.4\textwidth]{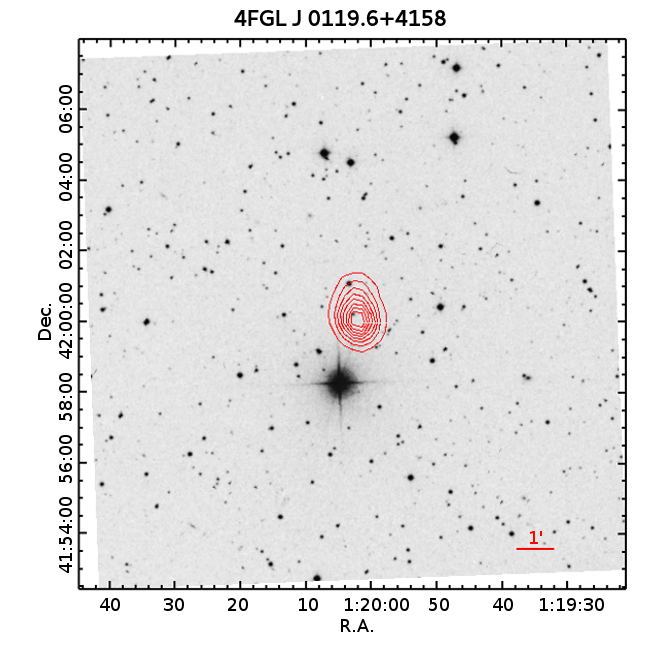} 
\includegraphics[width=.4\textwidth]{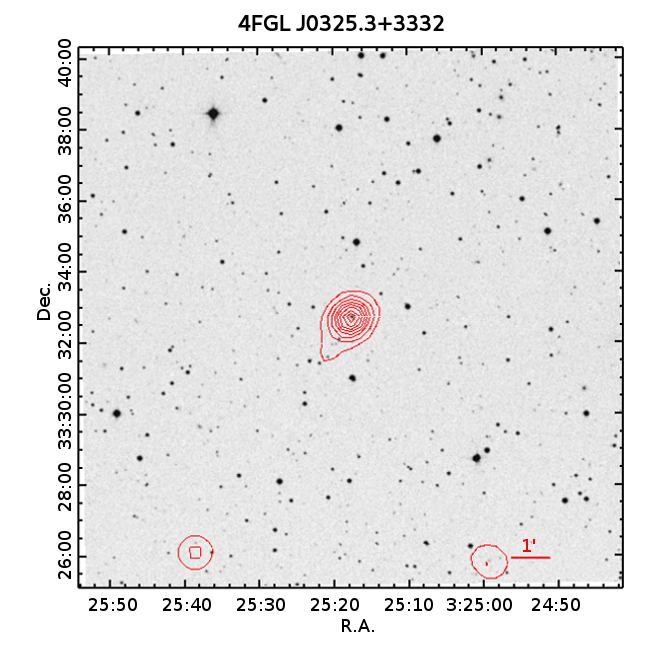} 
\includegraphics[width=.4\textwidth]{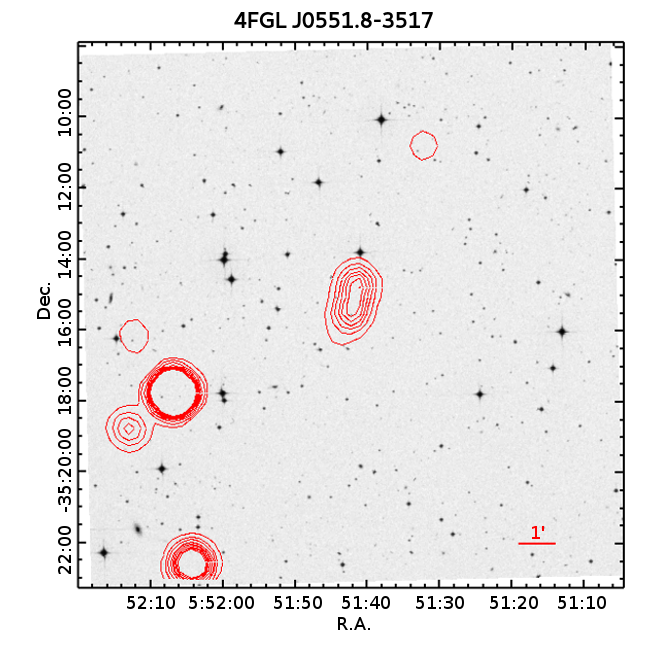} 
\includegraphics[width=.4\textwidth]{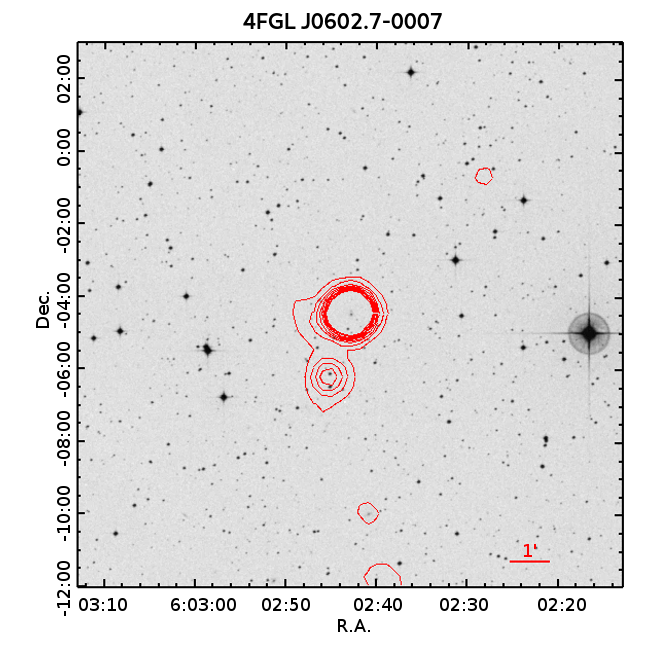} 
\includegraphics[width=.4\textwidth]{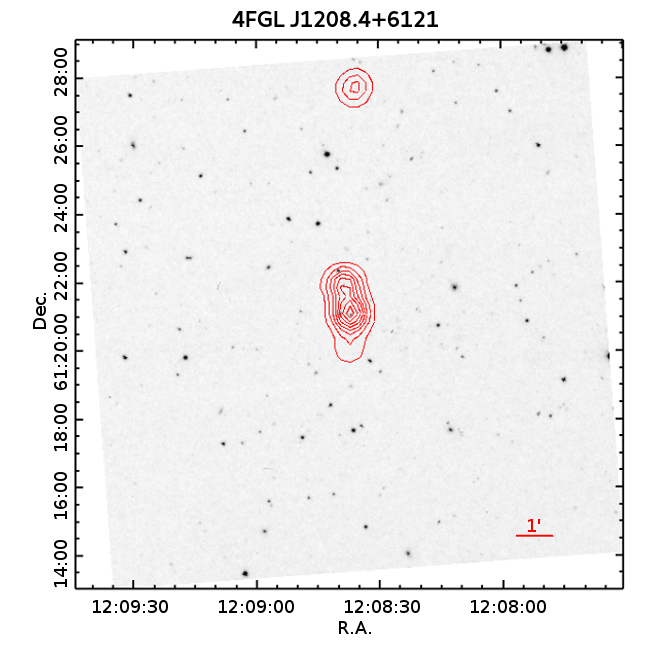} 
\includegraphics[width=.4\textwidth]{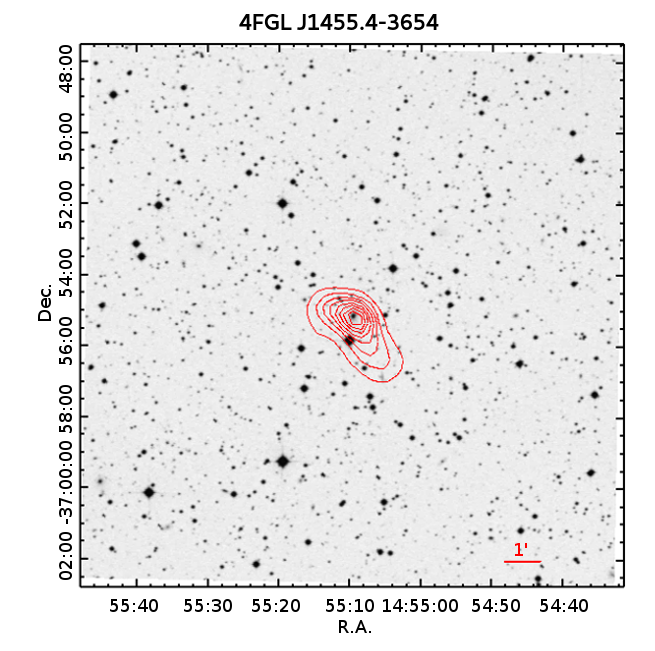} 
\includegraphics[width=.4\textwidth]{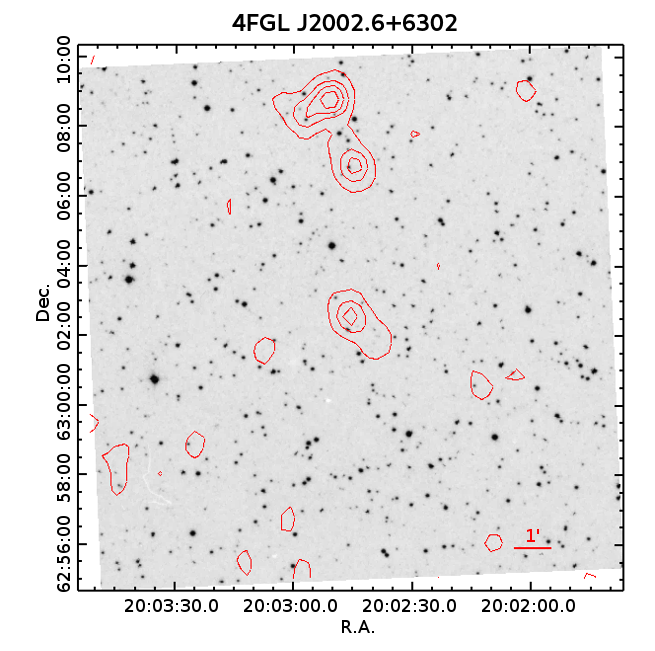}
\end{center}
\caption{Optical images and NVSS radio contour maps of the best MAGN candidates of the sample (see Table 2). In spite of the low survey observations resolution, all the candidates show an asymmetric extended component.
\label{<nvss>}}
\end{figure}

\section{Results}
\label{<3>}
\subsection{Properties of our MAGN candidates}
\begin{figure}
    \centering
    \includegraphics[width=0.9\textwidth]{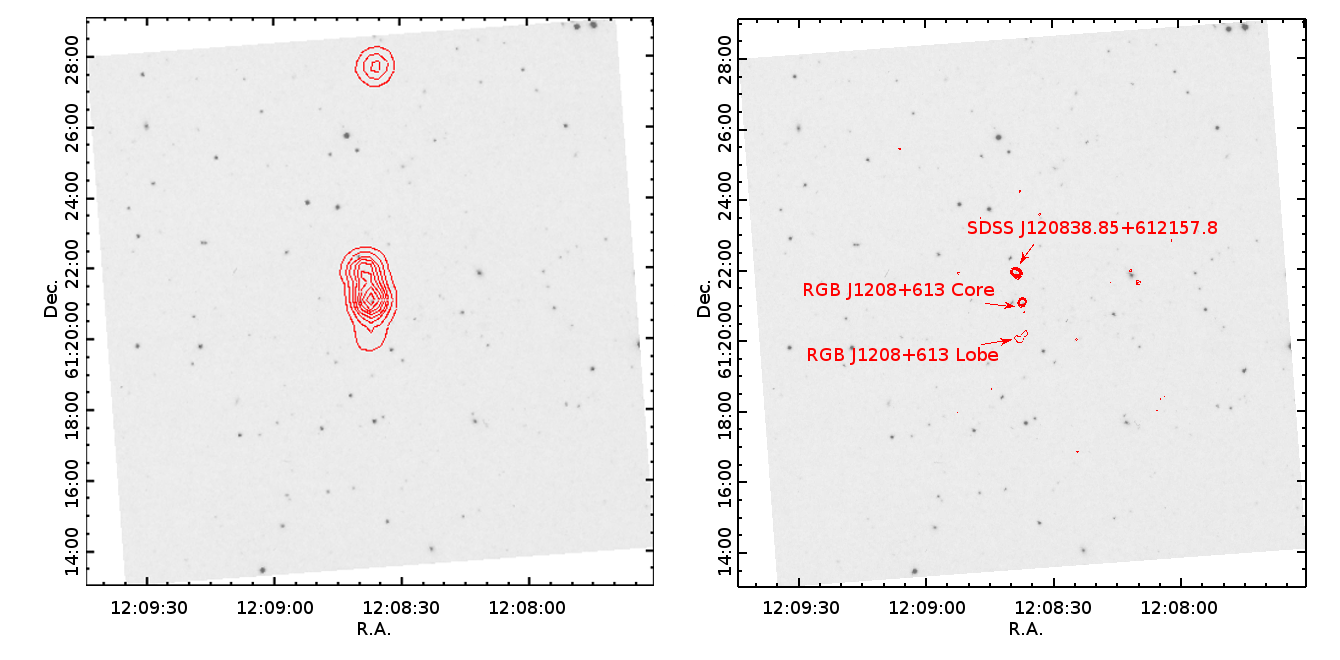}
    \caption{Comparison of the low resolution NVSS radio contour map of RGB J1208+613 (left panel) with the high resolution FIRST observation of the same source (right panel). \label{<first>}}
\end{figure}

The sources that have been identified in this study are all radio emitters with a clear detection in major all-sky radio surveys. The radio spectra tend to be rather flat or with indications of significant emission at high frequency, as inferred from the inclusion of these targets in lists of objects with flat radio spectra, such as the CLASS radio survey and the CRATES catalogue \citep{Myers01, Healey07}, or by their detection with high frequency observations, like the AT20G survey at $20\,{\rm GHz}$ \citep{Murphy10, Massardi10}. \\
Table~\ref{table2} presents the best MAGN candidates found in this study, presenting measurements of their angular size, together with additional information about their detection at X-ray frequencies, the availability of redshift and spectroscopic information and the corresponding projected sizes. Figure~\ref{<nvss>} shows radio contours plotted on top of optical images. These sources have radio images that show hints of partially resolved asymmetric structures, which point to the existence of an extended component. This interpretation can only be confirmed through the analysis of higher resolution radio images. The only case for which this test is possible, among our seven candidates, comes from the comparison of the low-resolution NVSS image and the high-resolution FIRST observation of RGB J1208+613 (4FGL J1208.4+6121), a galaxy located at $z = 0.275$. This comparison, illustrated in Figure~\ref{<first>}, reveals that the extended structure is produced by the presence of two bright hot spots, symmetrically distributed around the source. While the northern feature is probably associated with the emission of a nearby quasar, SDSS~J120838.85+612157.8 located at $z = 0.525$, the southern one appears as the termination of an extended emission, without other counterparts, that 
likely originated 
from RGB J1208+613 itself.

Although at the low resolution of the available survey images we are not able to infer the intrinsic structure of the radio sources, all the selected candidates feature an extended component. The emission is still core-dominated, but, unlike the case of blazars, which tend to be compact and symmetric, these objects show an extended structure that suggests hints of partially resolved morphology. These are more often asymmetric and one-sided, although in a smaller number of cases we can observe hints of two-sided emission. The core of the radio source is typically associated with a visible optical counter-part, which has faint or no detectable X-ray emission. Exceptions to the association of the radio sources with optical counterparts are generally occurring in objects that lie close to the Galactic plane, for which an effect of severe absorption prevents the observation of the visible light. The lack of a strong X-ray emission and the more abundant detection of asymmetric radio structures, over symmetric ones, are all indications of an intermediate inclination of the jets produced by these sources between the blazar and the radio galaxy regime. The spectra of the optical sources, when available, are also consistent with spectra of elliptical galaxies, characterized by strong absorption lines \citep{jon1, jon2, sdss1, sdss2, demen}.
These characteristics pose the sources belonging to our sample at an intermediate stage between closely aligned blazars and AGN with jets pointing at large angles relative to us, in agreement with 
our identification as new $\gamma$-ray MAGNs.

 \subsection{TeV candidates }
\begin{figure}
\label{<sed>}
\begin{center}
\includegraphics[width=.45\textwidth]{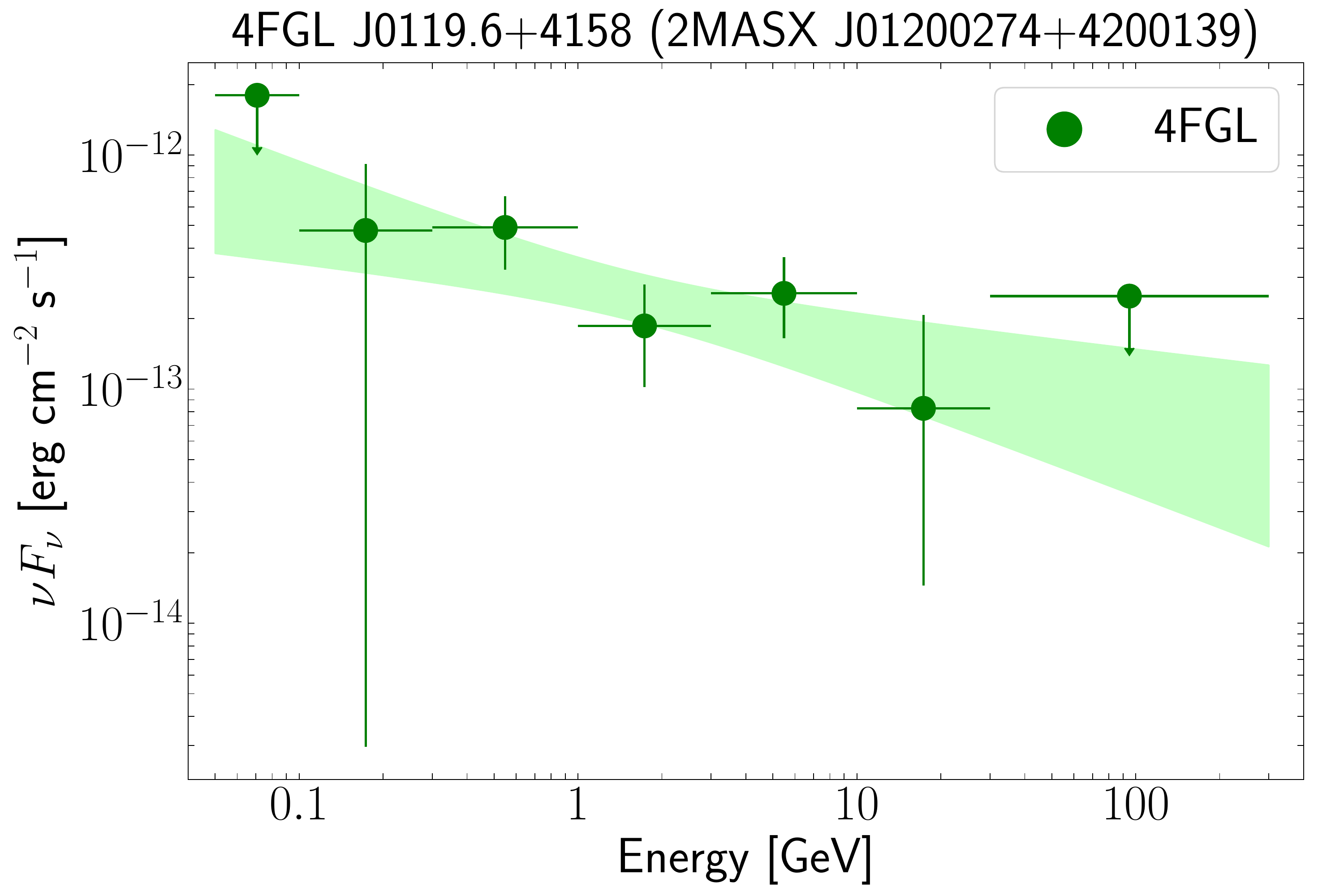}
\includegraphics[width=.45\textwidth]{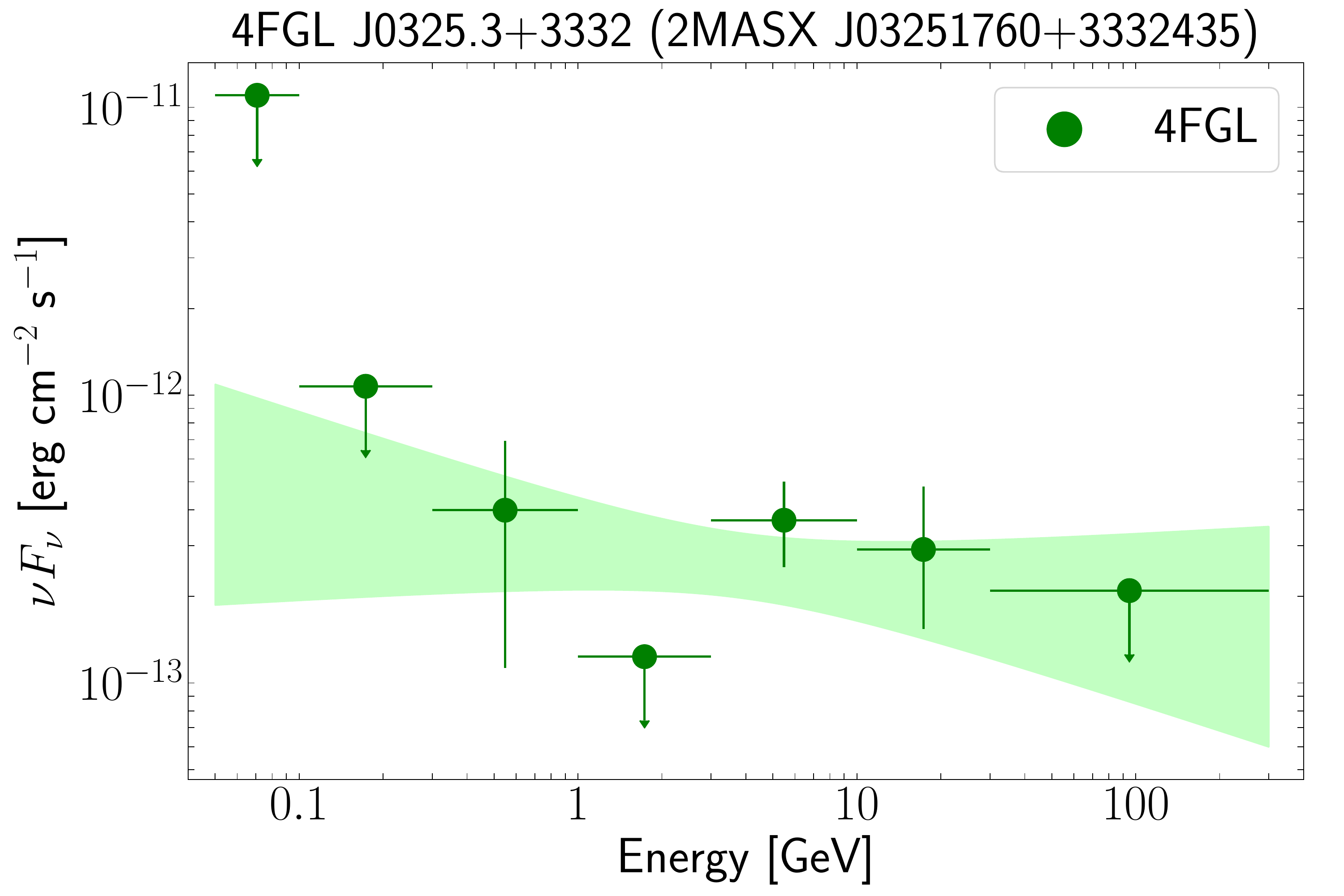}
\includegraphics[width=.45\textwidth]{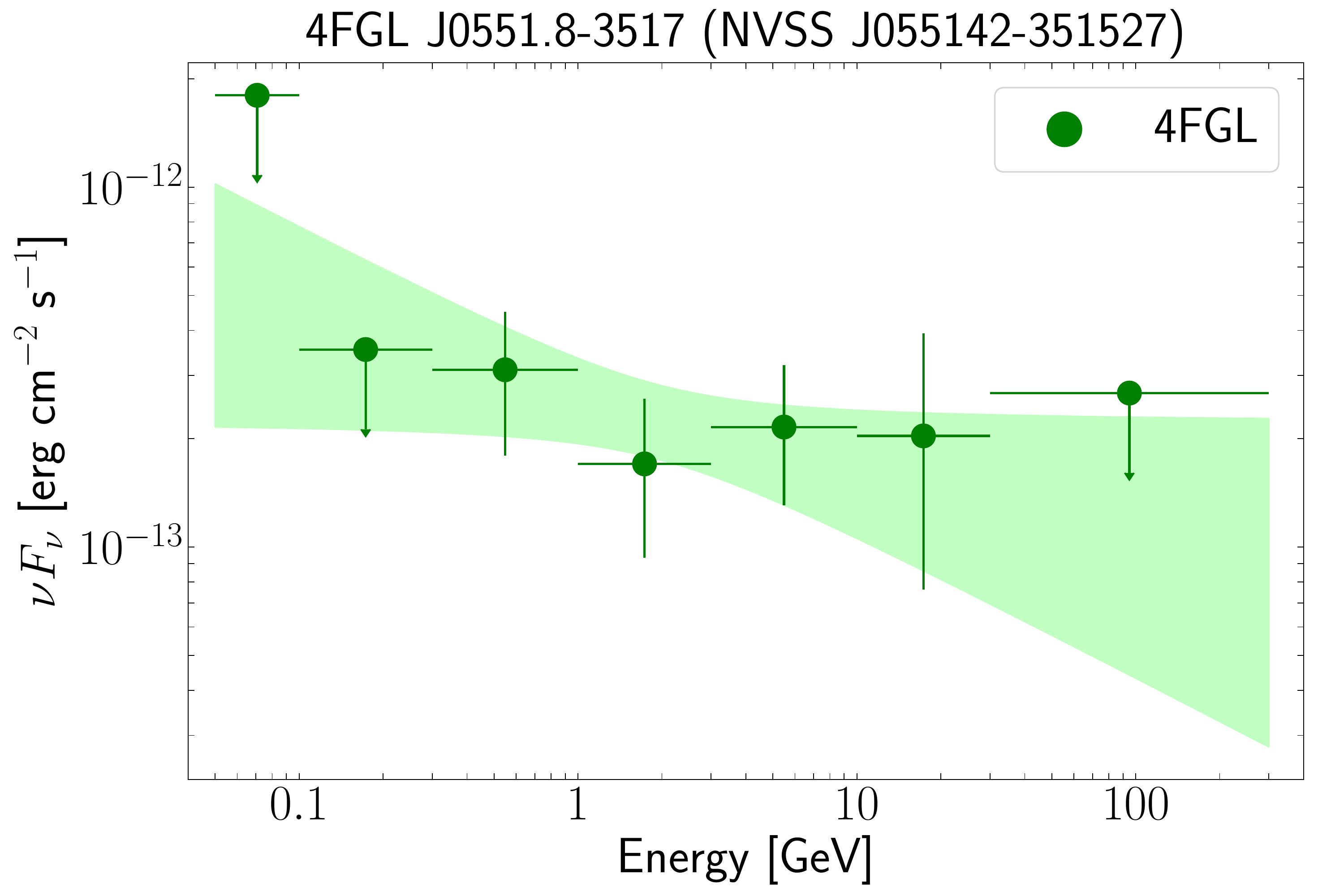} 
\includegraphics[width=.45\textwidth]{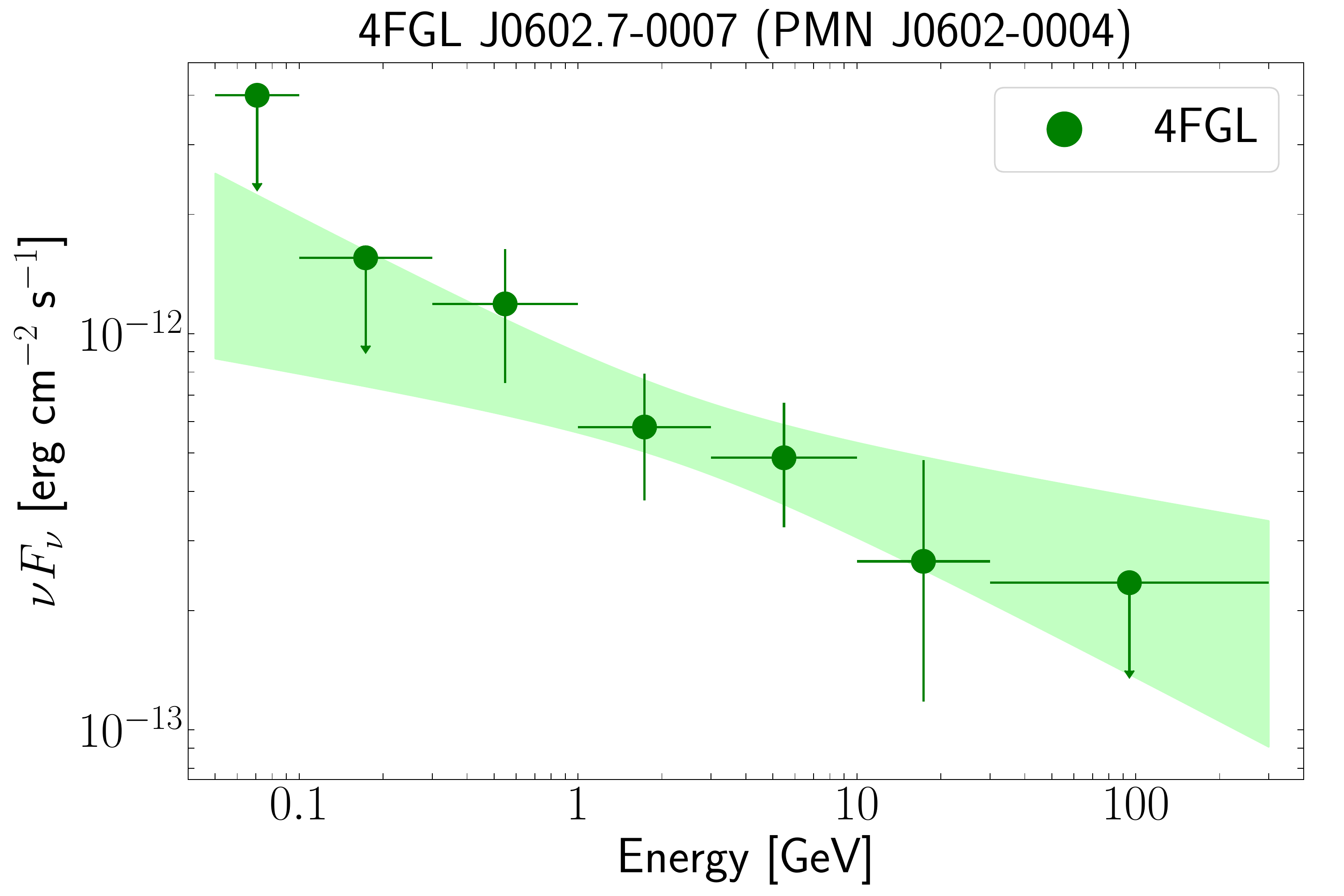}
\includegraphics[width=.45\textwidth]{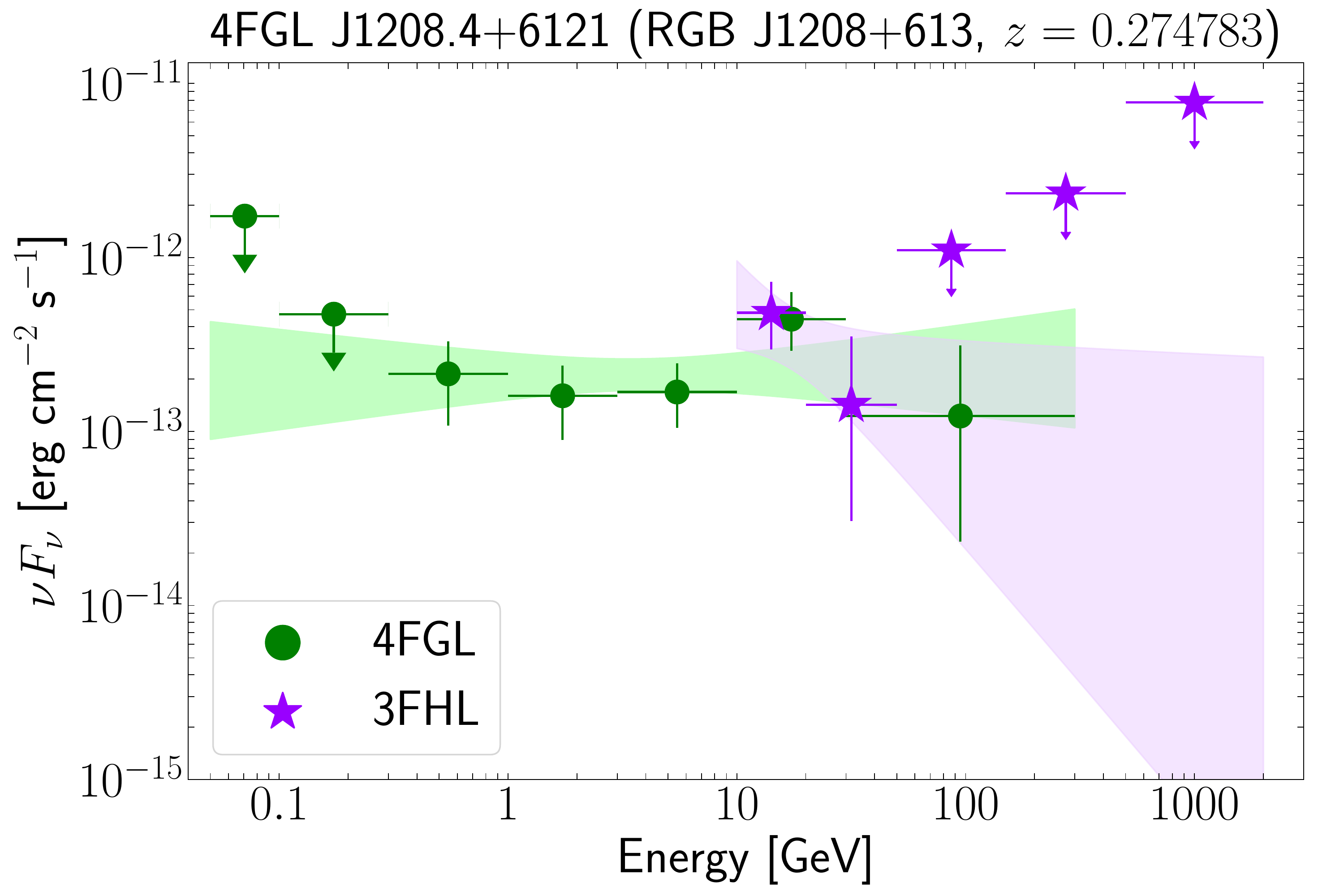}
\includegraphics[width=.45\textwidth]{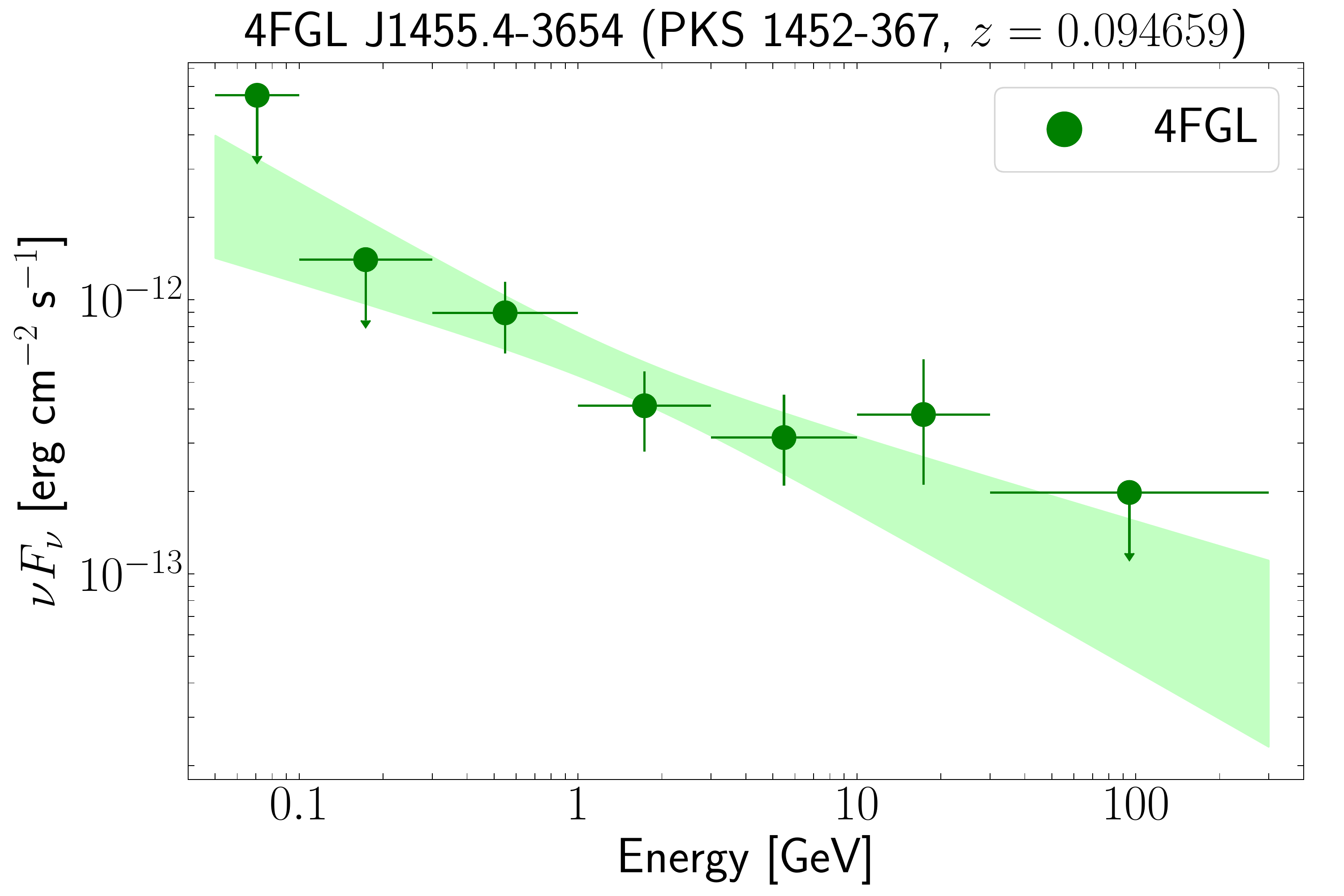}
\includegraphics[width=.45\textwidth]{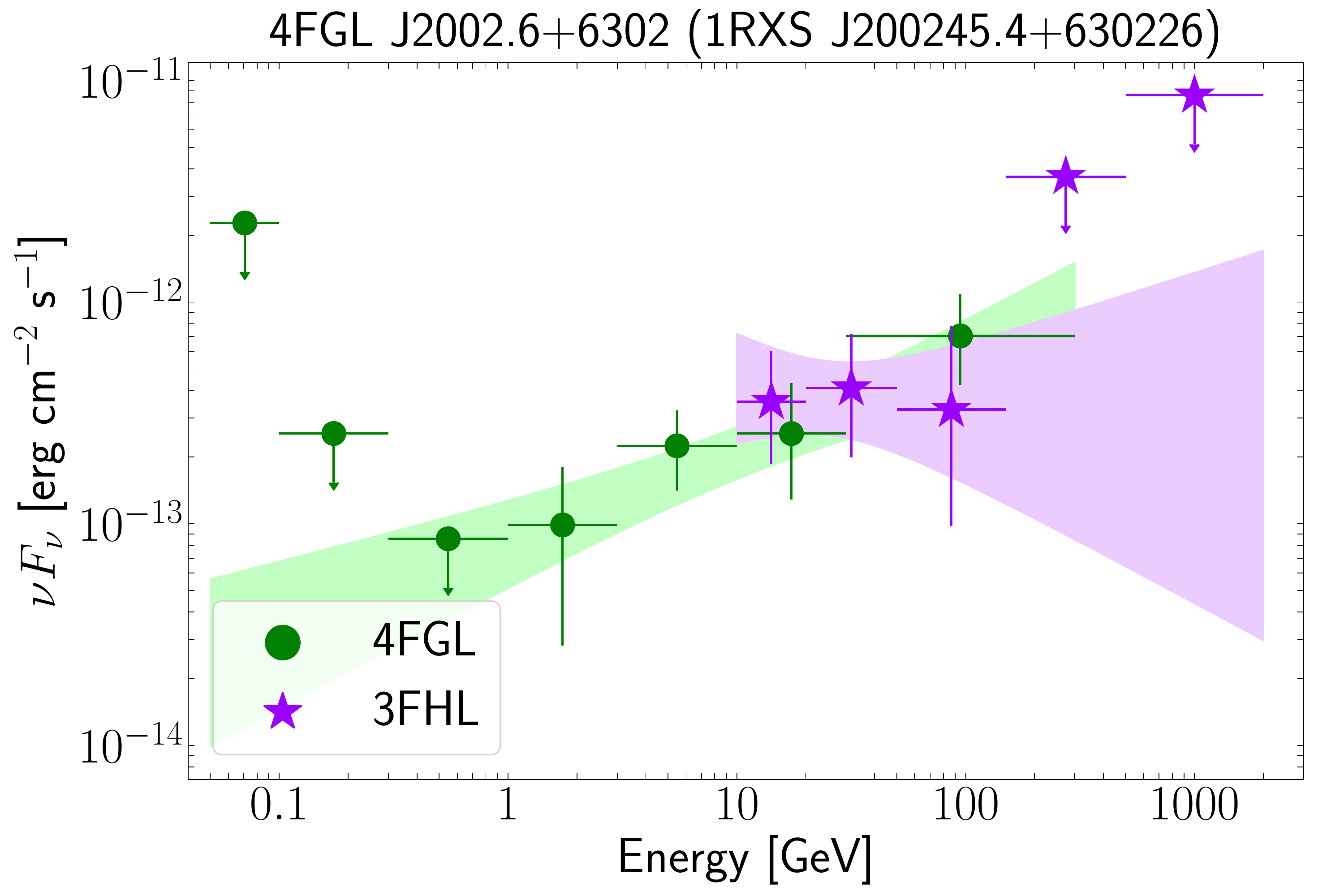}
\caption{Spectral data of our seven MAGN candidates from 4FGL (green) and 3FHL (violet). Note that the two catalogs are built using different exposures, therefore there may be variability issues between them. This figure shows that sources are faint and they can hardly be detected by IACTs.}
\end{center}
\end{figure}

We also checked the detectability at the TeV energy range for the MAGN candidates highlighted in Table~\ref{table2}. For this purpose we reviewed the $\gamma$-ray spectra of these sources, which are shown in Figure~\ref{<sed>}. We note that all of these sources are faint, with integrated energy fluxes from 100~MeV to 100~GeV lower than $8\times 10^{-12}$~erg/cm$^{2}$/s, and none of them has been detected by LAT at energies larger than about 30 GeV, except J2020.6+6302. For confirming this, we searched for these sources in the Third Catalog of Hard {\it Fermi}-LAT sources (3FHL, Ajello et al. 2017) and in the Second Catalog of Hard {\it Fermi}-LAT sources (2FHL, Ackermann et al. 2016). The former describes sources detected above 10 GeV in the first 7 years of mission, whereas the latter describes sources detected above 50 GeV in the first 6.7 years of mission. None of these sources are in 2FHL, and only two, J1208.4+6121 and J2002.6+6302, are in 3FHL. However, they are too faint to be detectable with Imaging Atmospheric Cherenkov Telescope (IACTs) in reasonable exposures unless they undergo a flare.

\section{Summary and conclusions}
\label{<4>}
We expanded the known population of 36 MAGNs currently classified in the 4LAC catalog by finding seven new candidates in the BCU population. Our method based on the morphological characterization of radio archive images and the study of the available optical observations identifies objects with strong evidence of MAGN nature, though only better resolution images can 
give conclusive evidence. Due to their radiative weakness, these sources cannot be observed in the TeV energy range.

\begin{table}
\caption{Properties of the best MAGN candidates. The table reports a measurement of the angular size of the extended emission, the redshift (if available), the corresponding scale, the projected size, a flag stating whether the source is detected in X-rays and the origin of the optical spectrum.}
\label{table2}
\begin{footnotesize}
\begin{tabular}{lcccccc}
\hline 
Source Name 	&	Ang. Size	&	{\it z}	&	Scale	&	Proj. Size	&	X-ray det.	&	Spec. Ref.$^{\rm a}$	\\
    & arcmin    &   &   ${\rm kpc} \cdot {\rm arcsec}^{-1}$ &   kpc &   &   \\
\hline		
4FGL J0119.6+4158	&	$1.41 \pm 0.20$	&	0.109	&	2.003	&	$169.4 \pm 24.0$	&	no	&	deM	\\
4FGL J0325.3+3332	&	$1.46 \pm 0.20$	&	0.128	&	2.303	&	$201.7 \pm 20.7$	&	no	&	deM	\\
4FGL J0551.8$-$3517	&	$1.50 \pm 0.15$	&	--	&	--	&	--	&	yes	&	--	\\
4FGL J0602.7$-$0007	&	$1.14 \pm 0.25$	&	0.118	&	2.147	&	$146.8 \pm 32.2$	&	yes	&	deM	\\
4FGL J1208.4+6121	&	$1.37 \pm 0.15$	&	0.274	&	4.215	&	$346.5 \pm 37.9$	&	yes	&	SDSS	\\
4FGL J1455.4$-$3654	&	$1.81 \pm 0.30$	&	0.095	&	1.774	&	$192.6 \pm 31.9$	&	yes	&	6dF	\\
4FGL J2002.6+6302	&	$1.41 \pm 0.25$	&	--	&	--	&	--	&	yes	&	--	\\
\hline
\end{tabular} \\
\end{footnotesize}
\footnotesize{$^{\rm a}$ deM = {\citet{demen}} \\ SDSS = Sloan Digital Sky Survey \\ 6dF = 6 degree Field Galaxy Redshift Survey}
\end{table}

\section{Acknowledgments}
\label{<5>}
Support for science analysis during the operation phase is gratefully acknowledged from the {\it Fermi}-LAT Collaboration for making the {\it Fermi}-LAT results available in such a useful form, the Institute of Space Astrophysics and Cosmic Physics in Milano - Italy (IASF), National Institute for Astrophysics (INAF) Rome - Italy  and  the Lab. de Instrumentacao e Fisica Experimental de Particulas. LIP, Lisboa, PT. A.D. acknowledges the support of the Ram{\'o}n y Cajal program from the Spanish MINECO. The authors would like the anonymous referee for discussion and suggestions leading to the improvement of this work. \\

\end{document}